\documentclass[preprint2]{aastex}

\newcommand{\ergcms}{ergs cm$^{-2}$ s$^{-1}$}
\newcommand{\hzs}{Hz s$^{-1}$}
\newcommand{\ergss}{ergs s$^{-1}$}

\begin{document}
\title{A Decade in the Life of EXO 2030+375: A Multi-wavelength Study of an
Accreting X-ray Pulsar}

\author{Colleen A. Wilson\altaffilmark{1}, Mark H. Finger\altaffilmark{2}}
\affil{SD 50 Space Science Research Center, National Space Science and 
Technology Center, 320 Sparkman Drive, Huntsville, AL 35805}
\email{colleen.wilson-hodge@msfc.nasa.gov}
\author{M.J. Coe, Silas Laycock}
\affil{Dept. of Physics and Astronomy, The University, Southampton, SO17 1BJ,
England}
\author{J. Fabregat}
\affil{Dept d'Astronomia i Astrofisica, Universitat de Valencia, 46100 
Burjassot, Valencia, Spain}

\altaffiltext{1}{NASA's Marshall Space Flight Center}
\altaffiltext{2}{Universities Space Research Association}

\begin{abstract}
Using BATSE and {\em RXTE} observations from 1991 April to 2001 August we have 
detected 71 outbursts from 82 periastron passages of EXO 2030+375, a 42-second 
transient X-ray pulsar with a Be star companion, including several outbursts 
from 1993 August to 1996 April when the source was previously believed to be 
quiescent. Combining BATSE, {\em RXTE}, and {\em EXOSAT} data we have derived 
an improved orbital solution. Applying this solution results in a smooth profile
for the spin-up rate during the giant outburst and results in evidence for a 
correlation between the spin-up rate and observed flux in the brighter BATSE 
outbursts. Infrared and H$\alpha$ measurements show a decline in the density of
the circumstellar disk around the Be star. This decline is followed by a sudden 
drop in the X-ray flux and a turn-over from a spin-up trend to spin-down in the
frequency history. This is the first Be/X-ray binary which shows an extended 
interval, about 2.5 years, where the global trend is spin-down, but the 
outbursts continue. In 1995 the orbital phase of EXO 2030+375's outbursts 
shifted from peaking about 6 days after periastron to peaking before periastron.
The outburst phase slowly recovered to peaking at about 2.5 days after 
periastron. We interpret this shift in orbital phase followed by a slow recovery
as evidence for a global one-armed oscillation propagating in the Be disk. This
is further supported by changes in the shape of the H$\alpha$ profile which are
commonly believed to be produced by a reconfiguration of the Be disk. The 
truncated viscous decretion disk model provides an explanation for the long
series of normal outbursts and the evidence for an accretion disk in the
brighter normal outbursts. Long-term multi-wavelength observations such as these
clearly add considerably to our knowledge of Be/X-ray binaries and the 
relationship between optical, infrared and X-ray observations.
\end{abstract}


\newpage
\section{Introduction}

Be/X-ray binaries are the most common type of accreting X-ray pulsar systems. 
They consist of a pulsar and a Be (or Oe) star, a main sequence star of spectral
type B (or O) that shows Balmer emission lines (See e.g., Slettebak 1988 and
Apparao 1994 for reviews.) The line emission is believed to be associated with an equatorial 
outflow of material expelled from the rapidly rotating Be star that probably 
forms a quasi-Keplerian disk near the Be star \citep{Quirrenbach97,Hanuschik96}.
X-ray outbursts are produced when the pulsar interacts with this disk. Be/X-ray
binaries typically show two types of outburst behavior: (a) giant outbursts (or
type II), characterized by high luminosities ($L_{\rm X} \gtrsim 10^{37}$
\ergss) 
and high spin-up rates  (i.e., a 
significant increase in pulse frequency) and (b) normal outbursts (or type I), 
characterized by lower luminosities ($L_{\rm X} \sim 10^{36}-10^{37}$ \ergss), 
low spin-up rates (if any), and recurrence
at the orbital period \citep{Stella86,Bildsten97}. As a population Be/X-ray 
binaries show a correlation between their spin and orbital periods
\citep{Corbet86,Waters89}. 

EXO 2030+375 is a 42-second transient accreting X-ray pulsar discovered during 
a giant outburst in 1985 with {\em EXOSAT} \citep{Parmar89}. Optical and 
infrared observations of the {\em EXOSAT} error circle identified a B0 Ve star 
as the most likely companion \citep{Motch87,Janot88,Coe88}. The initial outburst
was first detected at a 1-20 keV luminosity of $1 \times 10^{38}$ \ergss\ 
(using a distance of 5 kpc assumed by Parmar et al. 1989) on 1985 May 18 and 
declined to $\lesssim 3.8 \times 10^{34}$ \ergss\ by 1985 August 25.  During this
luminosity decline, the intrinsic spin period changed dramatically, with a 
characteristic spin-up timescale $-P/\dot P \approx 30$ yr \citep{Parmar89}. 
This large intrinsic spin-up suggested that an accretion disk was present and 
made determination of an orbit difficult, resulting in 3 acceptable orbits.  The
rate of change of pulse period $\dot P$ \citep{Parmar89}, the energy spectrum 
\citep{Reynolds93,Sun94} and the 1-10 keV pulse profile \citep{PWS89} 
all showed significant luminosity dependence. Further evidence of an accretion 
disk resulted from the detection of 0.2 Hz quasi-periodic oscillations 
\citep{Angelini89} consistent with the magnetospheric beat frequency model 
\citep{Lamb85} and the Keplerian frequency model \citep{vanderKlis87}. 
A second outburst, roughly a factor of 10 weaker in luminosity than the first,
was also detected with {\em EXOSAT} in 1985 October 28-November 3 \citep{Parmar89}.
This outburst was apparently a normal outburst, but it showed unusual flaring activity.
These flares were detected on 1985 October 30-31 and had a 4 hour recurrence
period \citep{Parmar89,Apparao91}. Observations of EXO 2030+375 were quite
sparse for the next several years. {\em Ginga} observed EXO 2030+375 on 1989 October
29-31 and 1991 October 24 \citep{Sun92}. Pulsations were detected only in the
1989 observations, with an observed pulse period of 41.68202(8) s and an
observed period derivative of $\dot P = -(8.3 \pm 0.9) \times 10^{-9}$ s/s at MJD
47828.95. EXO 2030+375 was also detected in the soft X-ray band for two days 
near a periastron passage with {\em ROSAT} in November 1990 \citep{Mavro94};
however the observations, short 10-28 s scans, were not suitable to detect
pulsations.

The most extensive observations of EXO 2030+375 were made with the Large Area
Detectors (LADs) of the Burst and
Transient Source Experiment \citep[BATSE]{Fishman89} on the {\em Compton Gamma Ray Observatory
(CGRO)}. From launch in April 1991 until {\em CGRO} was de-orbited in June 2000,
BATSE provided 
nearly continuous coverage of EXO 2030+375. During the interval 1992 February
8-1993 August 26, 13 consecutive outbursts of EXO 2030+375 were seen with
peak luminosities of $0.3 \times 10^{37} \leq L_{\rm X\ 1-20\ keV} \leq 3.0
\times 10^{37}$ \ergss\ (See Stollberg et al.\ 1999 for spectral assumptions), 
durations of 7-19 days, and spacings of 46 days 
\citep{Bildsten97,Stollberg99}. These outbursts peaked 5-6 days after periastron passage. A few 
detections of marginal statistical significance preceeded and followed this 
sequence of outbursts. During the 13 outbursts, the pulsar spun-up at an average
rate of $1.3 \times 10^{-13}$ \hzs. The 20-160 keV pulse profiles were double peaked
and showed no significant energy or luminosity dependence. A binary orbit listed
in the top row of Table~\ref{tab:orb} was determined using these 13 outbursts 
\citep{Stollberg99}. Near simultaneous optical, infrared, and X-ray observations of an outburst
in 1993 June/July showed no significant evidence for a correlation between X-ray
flux and infrared luminosity or between the X-ray flux and the equivalent width,
strength, or profile of the H$\alpha$ emission line \citep{Norton94}. EXO 
2030+375 was not detected above a flux level of $\sim 4.2 \times 10^{-10}$ 
\ergcms\ (20-50 keV) in the search 
techniques used from 1993 August 26 to 1996 April. Beginning in 1996 April, three short, weak,
outbursts were detected with BATSE, separated by 46 days 
\citep{Stollberg96,Stollberg99}. These outbursts began about 5 days prior to periastron passage as
predicted using the orbital model of \citet{Stollberg99}. Outbursts were also 
regularly detected with the All-Sky Monitor \citep[ASM]{Levine96} on the {\em Rossi X-ray Timing 
Explorer (RXTE)} beginning in 1996 March \citep{ReigCoe98}.

An outburst of EXO 2030+375 was observed from 1996 July 1-10 with the {\em RXTE}
Proportional Counter Array \citep[PCA]{Jahoda96}. Pulsations were detected throughout the
observations. This outburst began about 5 days prior to periastron passage.
The 2-10 keV pulse profile did not show significant intensity dependence. In
fact, it was consistent with the 1-10 keV {\em EXOSAT} profile observed at a 
similar luminosity \citep{ReigCoe98}. The energy spectrum was correlated with
luminosity \citep{Reig99}. This correlation was consistent with an extrapolation of that 
observed in {\em EXOSAT} data by \citet{Reynolds93}. Evidence for a possible cyclotron
feature at 36 keV was found in spectra from the High Energy X-ray Timing 
Experiment \citep[HEXTE]{Rothschild98} on {\em RXTE} \citep{Reig99}.  

In this paper we will present an improved orbit determination for EXO 2030+375
using BATSE, {\em RXTE}, and {\em EXOSAT} data. This improved orbit along with 
more sensitive search techniques has led to the detection of pulsations in 52 
outbursts in 9 years of BATSE data, including several outbursts in the period 
1993 August to 1996 April when the source was previously believed to be 
quiescent. Evidence for 19 additional outbursts, including 10 missed with BATSE
and 9 after {\em CGRO} was de-orbited, was observed with the {\em RXTE} ASM, for
a total of 71 outbursts observed out of 82 periastron passages from 1991 April
to 2001 August. We show that
our improved orbital parameters remove the two intervals of enhanced spin-up observed by 
\citet{Reynolds96} in the initial {\em EXOSAT} outburst, when they used the orbit of 
\citet{Stollberg94} to correct the observed pulse periods. We also show evidence
for a correlation between spin-up rate and flux in the BATSE data and compare it
to that observed in the {\em EXOSAT} data. We compare our X-ray results to
optical and infrared observations and discuss evidence for a decline in the
density of the Be 
disk and its effects on the X-ray flux and pulsed frequency histories. We 
discuss evidence in both optical observations of H$\alpha$ profiles and X-ray 
observations that suggests a global one-armed oscillation (i.e., a density 
perturbation) was propagating in the Be disk. Lastly, we discuss our
observations of EXO 2030+375 in context of current models.
 
\section{Analyses and Results}

\subsection{Discovery of Weak Outbursts with BATSE \label{sec:afs}}
In previous studies with BATSE \citep{Bildsten97, Stollberg99}, histories of 
pulse frequency and pulsed flux
for known pulsars were generated using grid searches over a range of candidate 
frequencies. The best fit frequency was determined using the $Z_{\rm n}^2$ 
statistic \citep{Buccheri83}. These studies were often limited to 1-day
integrations by systematic effects. We have developed an ``advanced" pulsar 
search that reduces 3 systematic effects: (1) aperiodic noise from sources in
the BATSE field of view, (2) Earth occultations of bright sources during the
folding interval, and (3) bright pulses from other pulsars in the BATSE field
of view. This technique is described 
in detail in \citet{Finger99} and \citet{WilsonH99}. We will summarize it here,
highlighting the differences from previous techniques.

The ``advanced" pulsar search technique consists of 3 steps (1) data selection 
and combination, (2) 20-50 keV pulse profile estimates, and (3) a grid search in 
frequency. First the BATSE
DISCLA channel 1 (20-50 keV, 1 s time resolution data) were selected for which
the source was visible, the high voltage was on, the spacecraft was outside the 
South Atlantic Anomaly, and no electron precipitation events or other anomalies
were flagged by the BATSE mission operations team. The count rates were combined
over the 4 LADs viewing EXO 2030+375 using weights optimized for an exponential
energy spectrum, $f(E) = A \exp(-E/kT)$ with temperature $kT = 20$ keV
\citep{Stollberg99}, and grouped into $\approx 300$ s segments. A segment length
of 300 s was used because it included several pulse periods, but was short 
enough that the background was well-fitted by a quadratic model.  The segment
boundaries were chosen to avoid Earth occultation steps from bright sources 
because these steps could produce spurious signals at high harmonics of the 
spacecraft orbital period which could overwhelm the real signal, especially for
long period pulsars. In the technique previously used by \citet{Bildsten97} and
\citet{Stollberg99} potentially bright sources were input manually, which worked
well for persistently bright sources, e.g., Crab or Cygnus X-1, but not as well
for transient sources. Effects from bright transients were greatly reduced in 
our advanced monitor by maintaining a database that included which sources were
active on a given day and what level they were active. Depending upon the pulse
period of the source, i.e., which spacecraft harmonic it was closest to, we 
automatically determined which occultation steps should avoided in the 
selection of segment boundaries. 

The second step in this process was estimation of an initial 20-50 keV pulse profile for 
each segment. In each segment, the combined rates were fitted with a model
consisting of a sixth-order Fourier expansion in pulse phase model 
(representing the 20-50 keV pulse profile) and a spline function with quadratics in time
(representing the background plus mean source count rate). A sixth-order Fourier
expansion was chosen based on the number of harmonics that are significant in a
1-day observation. Our initial pulse phase model was of the form $\phi(t^{\rm em}) = \nu_0
(t^{\rm em}-t_0)$, where $\nu_0 = 23.9954$ mHz was a constant barycentric 
frequency and $t^{\rm em}$ was the emission time corrected to the pulsar 
reference frame using the JPL DE-200 ephemeris \citep{Standish92} and the 
orbital parameters from \citet{Stollberg99}. The value and slope of the spline 
function were required to be continuous across adjacent segment boundaries, 
but not across data gaps. 

The final step in our advanced pulsar search was a grid search in frequency
using the set of typically several hundred estimated 20-50 keV pulse profiles from 4-day
intervals of data. New EXO 2030+375 pulse frequencies were determined from an 
initial grid search over 201 evenly spaced trial barycentric frequencies in the
range 23.97925-24.00240 mHz. Aperiodic noise from Cygnus X-1 caused the 
variances on the Fourier coefficients to be larger than expected for Poisson 
statistics. In previous searches \citep{Bildsten97}, this caused the $Z_{\rm n}^2$
statistic and BATSE's sensitivity to be dependent on the noise level since only
the Poisson level was assumed. \citet{Stollberg99} attempted to account for 
aperiodic noise from Cygnus X-1 using the average Cygnus X-1 noise level from a
single day; however, the noise level from Cygnus X-1 is highly variable 
\citep{Crary96}. Hence to better account for the aperiodic noise, its effect 
must be estimated for each frequency measurement. To correct the Poisson 
variances on the Fourier amplitudes at each harmonic\footnote{In this paper, 
harmonics are defined as $n\nu$ where $n = 1,2,3, \ldots$ and $\nu$ is the pulse
frequency.} for aperiodic noise in the power spectrum, we first computed the 
mean harmonic amplitudes, i.e., mean 20-50 keV pulse profile, for each 4-day interval.
Then we computed the reduced $\chi^2$ of a fit for each harmonic in the mean
20-50 keV pulse profile, to those from the 300-s segments within each 4-day interval. The
variances for each harmonic of the mean 20-50 keV pulse profile were then multiplied by 
this reduced $\chi^2$. Due to the large field-of-view of BATSE, other pulsars 
are often also present when we are measuring EXO 2030+375. If we limit our 
statistic to use the first 3 harmonics where EXO 2030+375 is the brightest, 
we reduce the chances of contamination of the search results from other pulsars
that happen to have harmonics near the higher harmonics of EXO 2030+375.
However, once we know the pulse frequency well and are no longer searching,
these harmonics can be re-included to improve the resolution of features in the
20-50 keV pulse profile. A modified $Z_3^2$ statistic, which we will call $Y_3$ after 
\citet{Finger99}, was then computed which incorporated the corrected variances. The reduced $\chi^2$ used 
to correct the variances on the first 2 Fourier amplitudes was $>2$ for the 
extended intervals MJD\footnote{MJD = Julian Date - 2400000.5} 48361-48650 
(1991 Apr - 1992 Jan), 49700-50000 (1994 Dec - 1995 Oct), 50400-50600 (1996 
Nov - 1997 Jun), 50650-50950 (1997 Jul - 1998 May), and several shorter 
intervals indicating increased noise from Cygnus X-1. The best-fit frequency 
for each 4-day interval was determined using the $Y_3$ statistic. The 
root-mean-squared (rms) pulsed flux was then estimated from the best-fit 20-50
keV pulse 
profile as \citep[Equation B9]{Bildsten97}
\begin{equation}
F_{\rm RMS} = \left[ 0.5 \sum_{k=1}^m (A_k^2 +B_k^2)
\right]^{1/2}\label{eqn:pflx}
\end{equation}
where $A_k$ and $B_k$ are the real and imaginary Fourier coefficients for each
harmonic. 

Figure~\ref{fig:freqs}, panels (a) and (b), illustrate the sensitivity 
improvements in our new techniques. Panel (a) shows the barycentered, orbit 
corrected pulse frequencies measured at 1-day intervals from \citet{Bildsten97}
plus a few later outbursts detected using a similar method. In that technique,
systematic effects dominated statistical errors for EXO 2030+375, resulting in
little improvement in sensitivity for 4-day intervals versus 1-day intervals. 
Panel (b) shows the barycentered, orbit corrected pulse frequencies 
measured at 4-day intervals from our new search. Accounting for systematic 
effects now allows us to combine 4-day intervals with near-statistic errors,
resulting in considerable improvement in sensitivity. 
Our new search revealed many more outbursts than had been previously detected, 
including several during the 2.5 year ``quiescent" interval \citep{Bildsten97} 
1993 August - 1996 April (MJD 49225-50175). However, there was considerable 
unexpected scatter in the pulse frequency measurements, starting during the 
so-called ``quiescent" interval. These outbursts were occurring at an earlier 
orbital phase than those used by \citet{Stollberg99} to determine the orbit.

\subsection{Orbit Fitting\label{sec:orb}}

We suspected that the scatter in the detected pulse frequencies might be 
produced by errors in the orbital parameters caused by coupling between the 
intrinsic spin variations of the pulsar with orbital effects.  The different 
orbital phase of the later outbursts gave us more orbital coverage, improving 
our ability to decouple these effects. One outburst in 1996 July 
(MJD 50265-50275) was also observed with the {\em RXTE} PCA 
\citep{ReigCoe98,Reig99}. Barycentered Standard 1 (125 ms, no energy resolution)
data were fitted with a model consisting of a constant background plus a 
sixth-order Fourier expansion in pulse phase model, creating an estimated 2-60
keV pulse profile for each PCA observation. The pulsed phase model consisted of a constant
barycentric frequency $\nu_0 = 23.9942$ estimated from BATSE measurements and
the orbital parameters of \citet{Stollberg99}. The rms pulsed fluxes where 
computed using Equation~\ref{eqn:pflx} for each 2-60 keV profile. The 2-60 keV 
pulse profile with
the brightest pulsed flux from MJD 50268.835-50268.867 (1996 July 4) was 
selected as the template profile. Phase offsets to the constant frequency model
were generated by cross-correlating the 2-60 keV pulse profiles from each observation 
with the template. Initially we suspected an error in the orbital period, which
would show up as an error in the projected epoch of periastron. To test this 
idea, we fitted the {\em RXTE} pulse phases with a quadratic phase model and an
orbit with the period fixed. The first and second rows in Table~\ref{tab:orb} 
list the \citet{Stollberg99} and our orbital parameters, respectively. 
Surprisingly, the epoch of periastron passage, $T_{\rm peri}$, was consistent 
(within 1 $\sigma$) with the value from \citet{Stollberg99} propagated to the 
epoch of the {\em RXTE} observations. However, the eccentricity $e$, the 
projected semi-major axis $a_x \sin i$, and the periapse angle $\omega$ all 
shifted slightly, by $2.2 \sigma$, $1.0 \sigma$, and $2.2 \sigma$, respectively.
Although these shifts were small, when we reran the frequency search using the 
revised orbital parameters, the frequency history no longer showed unexpected 
scatter. 

\begin{figure}
\plotone{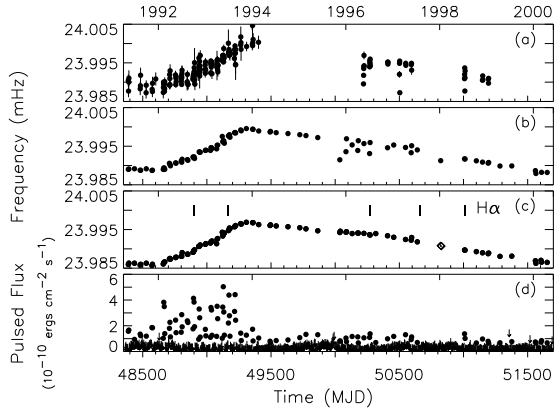}
\caption{Panels (a)-(c): Comparison of the sensitivity of two different 
frequency search techniques and two orbital solutions using BATSE data. Panel 
(a) shows detected pulse frequencies at 1-day intervals using techniques 
described in \protect\citet{Bildsten97}. Panels (b) and (c) show the 99.9\% 
confidence pulse frequency detections at 4-day intervals using our ``advanced" 
search technique that accounts for aperiodic noise from Cygnus X-1. The orbit of
\citet{Stollberg99} was used for panels (a) and (b), while panel (c) used our 
new orbital parameters in Table~\ref{tab:orb}, row 5, which considerably reduce
the scatter in the frequency measurements. An {\em RXTE} PCA frequency 
measurement is indicated by a diamond symbol. Vertical lines near the top of 
panel (c) denote times of H$\alpha$ measurements. Panel 
(d): Pulsed flux in the 20-50 keV band measured at 4-day intervals with BATSE. 
Upper limits (99\% confidence) are shown for all 4-day intervals where 
pulsations were not detected. 
\label{fig:freqs}}
\end{figure}

To better refine the orbital parameters, we generated pulse phase measurements 
at 1-day intervals using BATSE data for all outbursts detected in our initial
search. Table~\ref{tab:outbursts} lists the dates of the outbursts we 
detected. Following steps 1 and 2 of the advanced pulsar search (see
Section~\ref{sec:afs}), we generated estimated 20-50 keV pulse profiles for 300-s segments
using a pulse phase model consisting of a constant barycentric frequency for 
each outburst. For each 1-day interval, we computed a mean 20-50 keV pulse profile with
variances corrected for aperiodic noise using techniques described in
Section~\ref{sec:afs} and \citet{Finger99}. To better resolve features in the
20-50 keV pulse profiles, we retained all 6 harmonics. Phase offsets to the pulse phase 
model were generated by cross-correlating individual 20-50 keV pulse profiles with a 
template profile. The template profile was the best-fit 20-50 keV profile from the 
brightest 4-day interval (MJD 49126-49130, 1993 May 19-23) in our frequency
search. Pulsed fluxes in the 20-50 keV band were computed for each 1-day 
interval using Equation~\ref{eqn:pflx}. Previous studies 
\citep{Stollberg96,Bildsten97,Reig98} reported that the outburst observed in 
1996 April (MJD 50176-50183) occurred at an earlier orbital phase than the 13 
outbursts used by \citet{Stollberg99} to determine an orbit.  Comparing
Figures~\ref{fig:freqs} and \ref{fig:goao} (described in detail in
Section~\ref{sec:flx}), one sees that the outbursts
where the pulse frequencies show additional scatter, MJD 50000-51000 (1995 Oct -
1998 Jul), are occurring at a much earlier orbital phase than the 13 outbursts 
(MJD 48659-49228, 1992 Feb - 1993 Aug) used by \citet{Stollberg99} to determine
orbital parameters, suggesting that fitting these outbursts would result in 
improved orbital parameters. Pulse phase measurements from BATSE and {\em RXTE} PCA
were fitted with a global orbit plus a different quadratic for each outburst 
using the Levenberg-Marquardt method for $\chi^2$ minimization.  The best fit 
orbital parameters using the {\em RXTE} PCA data only, the {\em RXTE} PCA data 
plus the 13 outbursts used by \citet{Stollberg99}, and all 53 outbursts 
detected using the {\em RXTE} PCA and BATSE are listed in Table~\ref{tab:orb}. 
Although our pulse phase model does not fully describe the intrinsic pulse 
frequency variations, the different orbital phase coverage of the {\em RXTE} 
PCA observations and the later BATSE outbursts allow better decoupling of 
orbital and intrinsic torque effects.

Figure~\ref{fig:freqs}, panel (c), shows the barycentered and
orbit corrected (using Table~\ref{tab:orb} row 5) 4-day pulse frequency 
measured with BATSE ({\it filled circles}) and {\em RXTE} ({\it diamonds}). Comparing
the panels (b) and (c) of Figure~\ref{fig:freqs} indicates that the new orbital
parameters removed the scatter in the frequencies from MJD 50000-51000 (1995 
Oct - 1998 Jul). Also, in panel (c), there are 4 additional detected frequencies
above 99.9\% confidence. The bottom panel shows the 20-50 keV pulsed flux 
history measured with BATSE assuming an exponential energy spectrum with 
$kT = 20$ keV. Monte Carlo simulations were run to determine detection 
confidence levels. From $10^4$ trials, we determined detection confidence levels
of 99\% ($Y_3 \gtrsim 26.8$) and 99.9\% ($Y_3 \gtrsim 32.5$). In 
Figure~\ref{fig:freqs} (bottom panel), detections ($\gtrsim 99.9\%$ confidence)
are denoted using filled circles and 99\% confidence upper limits 
on the 20-50 keV pulsed flux for the periastron passages where BATSE did not 
detect pulsations from EXO 2030+375 are denoted by arrows. 

To further improve our orbital model, we decided to revisit the {\em EXOSAT}
observations of the initial giant outburst. We downloaded 1-8 keV background 
subtracted {\em EXOSAT} light curves from the Medium Energy (ME) proportional
counter \citep{Turner81} with 1 second time resolution from the High 
Energy Astrophysics Science Archive Research Center 
(HEASARC)\footnote{\url{http://heasarc.gsfc.nasa.gov}} at Goddard Space Flight 
Center for 14 {\em EXOSAT} observations 
\citep[for observation details]{Parmar89}. We were unable to unambiguously 
maintain a pulsar cycle count across these observations, so we measured pulse 
frequencies instead. For each observation, we fitted a model consisting of a 
constant plus a 6 harmonic Fourier expansion in a pulse phase model which
incorporated the orbit listed in the fourth row of Table~\ref{tab:orb}. The 
position history of the {\em EXOSAT} spacecraft versus time was not available, 
so we corrected to the solar system barycenter assuming a fixed spacecraft 
position. Effects of the spacecraft motion are expected to be small relative to
the measurement errors. We searched over a grid of 101 
evenly spaced pulse frequencies in the range 23.881-23.997 mHz. The best-fit 
frequency was selected using a method similar to that used with BATSE data. In 
this case, the aperiodic noise level was estimated by first extracting and then
averaging the Leahy normalized \citep{Leahy83} power spectra for all available 
1 ksec intervals of contiguous data in each observation. The variances on the 
$n-$harmonic amplitudes were multiplied by the ratio $\bar P_n/P_{\rm Poisson}$, 
where $\bar P_n$ is the average power in the frequency range $n\nu/2$ to 
$3n\nu/2$ and $P _{\rm Poisson}=2$ is assumed. The $Y_6$ statistic was then 
computed for each frequency in the grid for each observation. We confidently 
detected pulsations in all of the 14 observations except the observation on 
1985 August 25, when \citet{Parmar89} also failed to detect pulsations. 

\begin{figure}
\plotone{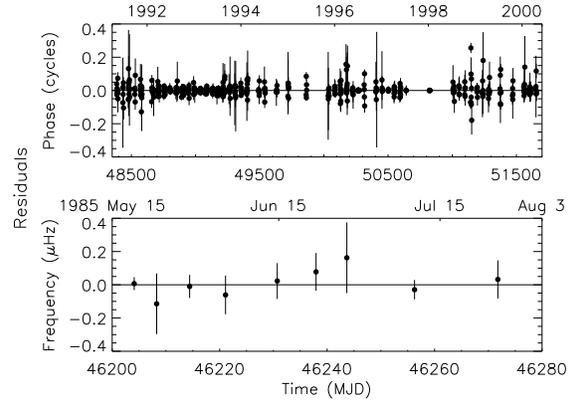}
\caption{Pulse phase residuals from BATSE and {\em RXTE} data (top panel) and 
frequency residuals from {\em EXOSAT} data (bottom panel) resulting from our \
joint fit to these data. The resulting orbital parameters are listed in 
Table~\ref{tab:orb}, row 5.
\label{fig:resid}}
\end{figure}

\citet{Reynolds96} corrected the {\em EXOSAT} pulse frequency measurements using the
orbital parameters of \citet{Stollberg94}. They reported that the spin-up trend 
in the intrinsic frequencies was not smooth, steepening temporarily between the 
1985 May 29  (MJD 46214) and June 4 (MJD 46220) observations, and again between
 July 10 (MJD 46256) and 25 (MJD 46271). They
speculated that this steepening could be due to errors in the orbital
parameters. When we used our new orbital parameters from BATSE and {\em RXTE}
data to correct the pulse frequencies, these changes in spin-up trend were 
reduced, but not completely removed. The {\em EXOSAT} data covered orbital 
phases not included in any BATSE or {\em RXTE} detections of EXO 2030+375. We 
decided to attempt an additional orbital fit to the combined data set, including
{\em EXOSAT} observed pulse frequencies from the initial giant outburst, BATSE 
pulse phases for 51 outbursts, and {\em RXTE} PCA pulse phases for 2 outbursts
(1996 July and 3 observations spanning 1998 January 8-15).
The fitting model consisted of a global orbit, a third order polynomial 
intrinsic pulse frequency model for the {\em EXOSAT} data, and a different 
quadratic pulse phase model for each outburst in the BATSE or {\em RXTE} PCA 
data. Table~\ref{tab:orb} (fifth row) lists the orbital parameters for the 
combined fit. Figure~\ref{fig:resid} shows the phase residuals (top panel) from
BATSE and {\em RXTE} and frequency residuals (bottom panel) from {\em EXOSAT} 
for this fit. 

\begin{figure}
\plotone{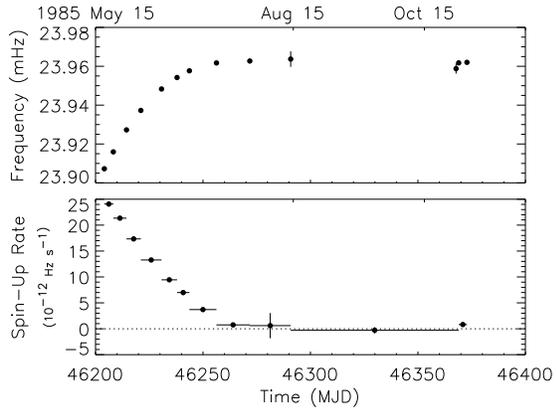}
\caption{Spin frequencies and spin-up rates measured using {\em EXOSAT}. The
top panel shows the barycentered pulse frequency corrected for the orbit in
row 5 of Table~\ref{tab:orb} for the 13 observations where pulsations were
detected. The bottom panel shows the spin-up rate computed by differencing
adjacent frequency measurements and dividing by the corresponding time
difference (denoted with horizontal lines). The spin-up rate measurement from 
1985 October 29-30 (MJD 46367-68) is not plotted because it had an error of 
$2.4 \times 10^{-11}$ \hzs, due to the short duration of the October 29 (MJD
46367) observation ($< 3000$ seconds) and due to the short, 1.2 day, spacing 
between the two observations. 
\label{fig:exosat}}
\end{figure}

Figure~\ref{fig:exosat} shows the intrinsic 
pulse frequencies from the outbursts observed with {\em EXOSAT}, barycentered 
and corrected for the orbit from the combined fit. The bottom panel shows the spin-up
rate computed by differencing adjacent pulse frequencies. The pulse frequency 
measurement on 1985 October 29 (MJD 46367) was omitted because it had a large 
measurement error due to the short $< 3000$ second observation. At the end of 
the giant outburst and in between the outbursts, the pulse frequencies are 
consistent with a constant value. During the second outburst, the average 
frequency derivative was $(9.7 \pm 3.9) \times 10^{-13}$ \hzs, suggesting mild 
spin-up, similar to many of the weaker outbursts observed with BATSE and 
{\em RXTE}. The spin-up rate during the giant outburst is now consistent with a 
smooth trend. The improved orbital parameters derived from the joint fit to
{\em EXOSAT}, BATSE, and {\em RXTE} data have removed the temporary enhanced spin-up
reported by \citet{Reynolds96}. 

Figure~\ref{fig:longterm} shows the long-term history of the spin frequency of 
EXO 2030+375, barycentered and corrected using the orbit from this joint fit.
This history includes spin frequency measurements from the {\em EXOSAT}, BATSE,
and {\em RXTE} data used in our orbital fits. In addition, the published pulse
frequency measurement from 1989 {\em Ginga} observations \citep{Sun92} is
shown and has been corrected using our improved orbital parameters. The {\em
Ginga} observations were not used in our orbital analysis because they spanned
less than 3 days, which was too short to extract additional orbital information.
During the initial giant outburst observed with {\em EXOSAT}, EXO 2030+375 
spun-up at an average rate of $(1.1 \pm 0.1) \times 10^{-11}$ \hzs. Between the
second outburst observed with {\em EXOSAT} and the first outburst observed with
BATSE, the average spin-up rate was two orders of magnitude smaller than in the
giant outburst, but still substantial at $(1.37 \pm 0.07) \times 10^{-13}$ \hzs.
During the BATSE era, EXO 2030+375 exhibited 
both spin-up and spin-down. From {\em CGRO} launch until about 1992 February 
(MJD 48660), EXO 2030+375 was consistent with a constant spin frequency. For 
about 1.8 years from 1992 February until 1993 November (MJD 48660-49308), the 
average spin-up rate was $(1.91 \pm 0.04) \times 10^{-13}$ \hzs, during the 
brighter outbursts observed with BATSE. When the outbursts became fainter, EXO 
2030+375 began to spin down at an average rate of $(-5.3 \pm 0.1) \times 
10^{-14}$ \hzs, for the last $\sim 6.5$ years of the BATSE mission. The average
spin-up rates suggest that EXO 2030+375's spin history was different during the intervening period 
between {\em EXOSAT} and BATSE observations than during the BATSE era. Between 
the last {\em EXOSAT} and first BATSE observation, separated by about 5.5 years,
there is significant average spin-up, suggesting that the pulsar was spinning-up
for a significant fraction of the time. In contrast, the average spin-up rate
during the first 5.5 years of the BATSE mission was a factor of $\sim 2$
smaller. Another giant outburst with a frequency change equal to that in the 
{\em EXOSAT} outburst was probably not missed. If one occurred just after the 
second outburst observed with {\em EXOSAT}, a spin-down rate of $\gtrsim 4$ 
times that observed with BATSE would be needed to spin the pulsar down to the 
spin frequency observed with BATSE. Interpretation of source behavior based on
average spin-up rates needs to be treated with caution.  Comparing only the 
spin frequencies from the first and last BATSE observation, to mimic the sparse 
observations prior to the BATSE era, one would erroneously conclude that EXO 
2030+375 had remained at an approximately constant spin frequency over 9 years, which is clearly not the case. 

\subsection{Spin-up Torque vs. Flux correlations\label{sec:torq}}

If an accretion disk is present in an accreting pulsar system, we expect to see
a correlation between spin-up and flux based on accretion theory. In the giant
outburst observed with {\em EXOSAT} such a correlation was observed
\citep{Parmar89, Reynolds96, Stollberg99}. The detection of a QPO at 
0.2 Hz by \citet{Angelini89} provided further evidence of an accretion 
disk. However, previous studies \citep{Stollberg99} were
unable to determine if such a correlation was also present in the BATSE data due
to coupling between the torque model and orbital parameters.  Since we now have
a good orbital solution that is not strongly coupled to the assumed torque 
model, we can address whether or not the spin-up rate is correlated with flux in
the BATSE observations. Here we first consider the BATSE data, then revisit 
{\em EXOSAT} data, and lastly estimate a bolometric correction to 
compare them. 

\begin{figure}
\plotone{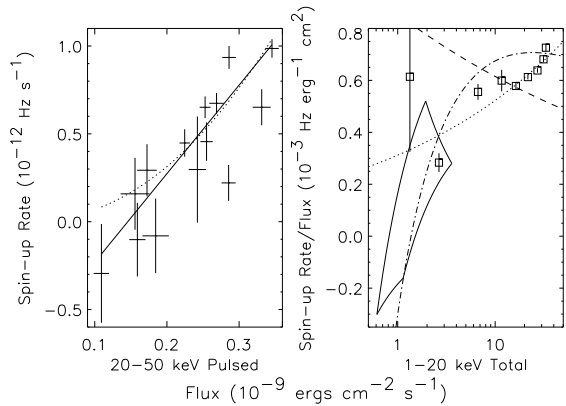}
\caption{(left-hand panel): The average spin-up rates versus the average 20-50 
keV pulsed fluxes on a linear scale for the brighter outbursts observed with 
BATSE. The solid and dotted lines are the best-fit linear and power law models,
respectively, to the BATSE data. 
(right-hand panel): {\em EXOSAT} spin-up rates divided by the 1-20 keV total 
flux vs. the 1-20 keV total flux. The point with the largest error bar is from
the normal outburst observed with {\em EXOSAT}, while the rest are from the
giant outburst. Our best-fit power law, with an index of 1.17, and a power law 
with an fixed index of $6/7$ are denoted by dotted and dashed lines, 
respectively. The dot-dashed line is the \citep{Ghosh79} model fitted to
the {\em EXOSAT} data by \citep{Reynolds96}. The solid box denotes the
projection of the linear fit to the BATSE data assuming a range of bolometric
corrections from 5.6-10.4, based on our estimated correction of 8.0 with an
assumed systematic error of 30\%.  
\label{fig:fdotvsflux}}
\end{figure}

For each outburst observed with BATSE, we fitted the one day pulse phase 
measurements with corresponding barycentered and orbit corrected arrival times 
using a quadratic phase model. This model gave us the average spin-up rate for
each outburst. We averaged the 1-day pulsed flux measurements for each outburst
to get the average 20-50 keV pulsed flux. Figure~\ref{fig:fdotvsflux} (left 
panel) is a linear plot of the BATSE spin-up rates versus 20-50 keV pulsed flux.
The BATSE spin-up rates are correlated with the BATSE pulsed fluxes, with a
linear correlation coefficient of 0.85 with a chance probability of $10^{-4}$. 
We fitted the data with a power-law (dotted line in the left panel of 
Figure~\ref{fig:fdotvsflux}), obtaining an index of $2.2 \pm 0.6$, although the
fit was formally unacceptable. We also fitted the BATSE data with a linear 
model. The best-fit line to the BATSE data (solid line in 
Figure~\ref{fig:fdotvsflux}) had a slope of $(5.0 \pm 1.0) \times 10^{-3}$ Hz 
erg$^{-1}$ cm$^2$ and an x-intercept (20-50 keV pulsed flux at zero spin-up) of
$(1.5 \pm 0.3) \times 10^{-10}$ \ergcms. 

Figure~\ref{fig:fdotvsflux}, right-hand panel, shows the spin-up rates divided 
by the 1-20 keV flux versus the 1-20 keV flux from {\em EXOSAT} observations. 
These units are chosen to better illustrate deviations from simple power law and
the \citet{Ghosh79} models than a log-log plot of spin-up rate versus flux. 
Spin-up rates (shown in Figure~\ref{fig:exosat}), were computed by differencing 
adjacent spin frequencies measured using our improved orbital parameters. Fluxes
in the 1-20 keV band were computed from the luminosities in Table~1 in 
\citet{Parmar89}. The 1-20 keV flux corresponding to each spin-up rate was 
computed by averaging adjacent fluxes. We fitted the {\em EXOSAT} data with a 
power law (dotted line in right panel of Figure~\ref{fig:fdotvsflux}), obtaining
an index of $1.17 \pm 0.04$ (denoted by a dotted line), although the fit was 
formally unacceptable. Previous fits to these data using different orbital 
solutions gave power law slopes of 1.1-1.4 \citep{Parmar89}, 1.2 
\citep{Reynolds96}, and $1.02 \pm 0.12$ \citep{Stollberg99}, although all of
these fits were also formally unacceptable. In the right-hand 
panel of Figure~\ref{fig:fdotvsflux}, a dashed line denotes our best-fit power 
law with an index fixed at $6/7$ which represents the relationship between 
spin-up and flux predicted by simple accretion theory. We have also overlaid 
\citet{Reynolds96}'s fit of the \citet{Ghosh79} model as a dot-dashed line. (See
Reynolds et al. 1996 for the model formulation and parameters).  

To compare the BATSE and {\em EXOSAT} data, we computed an approximate
bolometric correction using the faint 1996 July outburst which was observed by
both BATSE and {\em RXTE}. The average 20-50 keV pulsed flux for MJD 50266-50274
(1996 July 2-10) was $1.18 \times 10^{-10}$ \ergcms\ and the average 2.7-30 keV 
flux measured with {\em RXTE} for those dates was $9.47 \times 10^{-10}$ 
\ergcms\ \citep{Reig99}. Taking the ratio of the two fluxes gives a bolometric 
correction of $\sim 8.0$ for the BATSE data. Several sources of systematic error
affect this correction including differences in energy calibrations of BATSE,
{\em RXTE}, and {\em EXOSAT}, and variations of the spectrum and pulse fraction
with luminosity. Because of these systematic errors, the comparison of the two
data sets can only be qualitative. The solid box in the right panel of 
Figure~\ref{fig:fdotvsflux} denotes the linear fit to the BATSE data projected 
on to this plot by assuming a range of bolometric corrections from 5.6-10.4, 
i.e., our estimated correction with an assumed systematic error of 30\%.

\subsection{X-ray Flux Measurements\label{sec:flx}}

\begin{figure}
\plotone{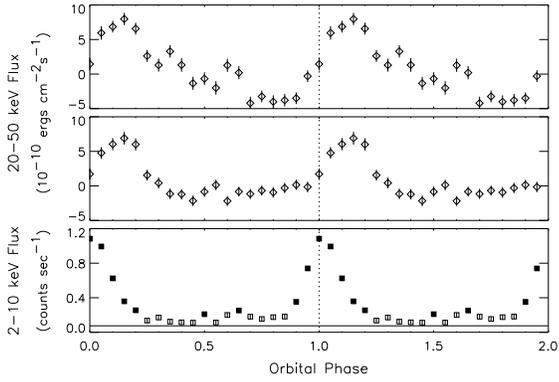}
\caption{(Top 2 Panels): Epoch-folded profiles of BATSE Earth occultation data from MJD
48363-49353 (1991 April - 1993 December). A period of 46.0214 days was used with
a periastron epoch of MJD 50547.22. The top panel includes rising occultations 
only and the center panel includes only setting occultations. The negative 
offset in flux is due to interference from primarily Cyg X-1 and Cyg X-3.
(Bottom Panel): Epoch-folded profiles of {\em RXTE} ASM measurements of 2-10 keV
flux from MJD 50135-52138 (1996 February - 2001 August). A period of 46.0214 
days was used with a periastron epoch of MJD 50547.22. The horizontal line 
denotes the 1 mCrab (0.075 counts s$^{-1}$) positive bias seen in long-term 
averages of ASM data. The orbital phase of periastron is denoted with a dotted 
vertical line in all 3 panels.\label{fig:epfold}}
\end{figure}

\begin{figure}
\epsscale{0.7}
\plotone{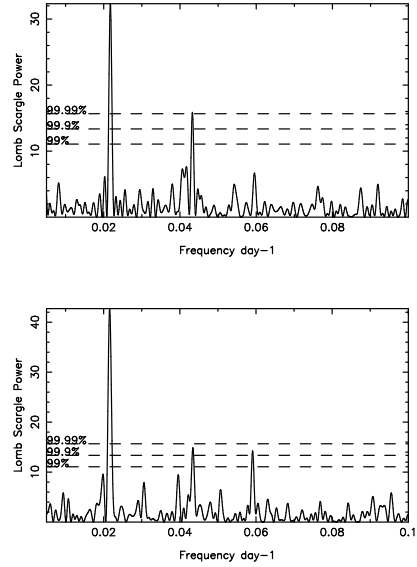}
\caption{Lomb-Scargle periodograms of BATSE Earth occultation data from MJD
48363-49353 (1991 April - 1993 December). The top panel includes rising occultations only and the bottom
panel includes only setting occultations. The confidence levels shown have
been determined by running multiple sets of randomized data with the same window
function as the BATSE data and the same statistical properties. The highest peak
in each panel is near the 100\% confidence level. The second highest peak is at 99.99\% and 99.95\% confidence in
the top and bottom panels respectively. 
\label{fig:ls}}
\end{figure}

BATSE provided nearly continuous monitoring of the whole sky in the 20 keV to 2
MeV band using the Earth occultation technique \citep{Wilson00, Harmon01}. When
a source went behind (or emerges from behind) the Earth, step-like features were
produced in the BATSE data twice every 90 minute orbit. To measure the intensity
of a known source, about 2 minutes of data (in each of 16 energy channels and 
each detector, using the 2.048 s CONT data) immediately before and after the 
occultation step were fitted with a
model consisting of a quadratic background plus source terms for the source of
interest and any interfering sources within 70\arcdeg\ of the detector normal.
Unfortunately, in the case of EXO 2030+375, the bright, highly variable sources
Cyg X-1 and Cyg X-3 produce systematic errors in the step measurements. When
these sources were not Earth occulted, they could produce rapid variations in the 
total count rate which could cause the size of the EXO 2030+375's steps to be 
under or overestimated. Variations in the bright source could also produce a
positive or negative offset in the average step measurement (See Harmon et al.
2001 for a detailed description of systematic errors). The strength of these 
systematic errors varied with both the brightness of the sources and the 52-day
precession period of the spacecraft. To attempt to address these systematic 
errors, we epoch-folded rising and setting occultation steps separately for MJD
48363-49535 (1991 Apr- 1993 Dec) when EXO 2030+375 was brighter, shown 
in the top and center panels, respectively, of Figure~\ref{fig:epfold}. Since 
rising and setting steps sampled different slices of the sky, features present 
in both plots are likely due to EXO 2030+375. However, the setting steps were 
less contaminated by Cygnus X-1 than the rising steps because Cygnus X-1 always
rose before EXO 2030+375, hence it always affected the rising step measurements,
but Cygnus X-1 also always set before EXO 2030+375. The setting steps from 
Cygnus X-1 were still within the fitting window for EXO 2030+375, but since 
Cygnus X-1 was active for less of the fitting window than in the rising steps,
it had a smaller effect. Evidence of these systematic effects can be seen by 
comparing the minima in the top and center panels of Figure~\ref{fig:epfold}. 
Systematic effects caused the minimum to be offset from zero by a negative value.
In the top panel, the minimum is about twice as negative as in the center panel,
suggesting more interference.  In both panels, there is a suggestion of 
emission near apastron. Further possible evidence for apastron emission is shown
in Figure~\ref{fig:ls} which shows Lomb-Scargle periodograms for the rising (top
panel) and setting (bottom panel) occultations from MJD 48363-49353 (1991 Apr -
1993 Dec). In both periodograms, there are significant peaks at the EXO
2030+375 orbital period $P_{\rm orb}$ (chance probabilities $8.1 \times
10^{-12}$ and $3.6 \times 10^{-16}$ for the top and bottom panels, respectively)
and at $P_{\rm orb}/2$ (chance probabilities $7.2 \times 10^{-5}$ and $5.3
\times 10^{-4}$ for the top and bottom panels, respectively). The chance 
probabilities and confidence levels (listed in the caption to Figure~\ref{fig:ls}) were computed using a 
Monte-Carlo simulation to generate the probability distribution for the null 
hypothesis over the frequency range $0.005-0.01$ d$^{-1}$ and then fitting 
Equation 13.8.7 from \citep{Press92} by least squares. These peaks are 
consistent with a double-peaked  orbital light curve. \citet{Reig98} reported 
$\sim 3\sigma$ evidence for apastron emission in {\em RXTE} ASM data from a 
later epoch, 1996-1998 February (MJD 50135-50870). To look for additional
evidence of apastron outbursts, we epoch-folded the available {\em RXTE} ASM 
1-dwell data, using a period of 46.0214 days and a periastron epoch of MJD 
50547.22. Figure~\ref{fig:epfold} (bottom panel) shows the folded profile from MJD 
50135-52138 (1996 February - 2001 August). Filled squares denote points that are $>5\sigma$ above
the 0.075 counts sec$^{-1}$ (1 mCrab) positive bias found in long-term averages of ASM data according to the ASM Instrument Team and
apparent in the sum band light curves of sources believed to be well below the
ASM threshold (e.g., SS 0019+21, X 0620--003, and PSR J1022+1001). This ASM folded profile shows $5.2 
\sigma$ evidence for apastron emission.

\begin{figure}
\plotone{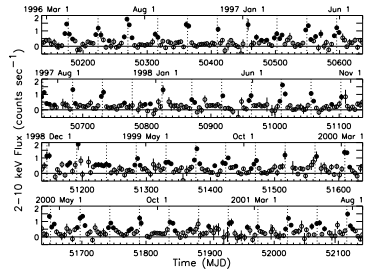}
\caption{
2-10 keV flux measured with the {\em RXTE} ASM. The
ASM measurements have been averaged over 4-day intervals. Filled circles denote
$\gtrsim 3 \sigma$ detections. Vertical dotted lines denote periastron passage
times.
\label{fig:asm}}
\end{figure}

Because EXO 2030+375 is brighter in the 2-10 keV band than in BATSE's 20-50 keV
band, the {\em RXTE} ASM more easily detects individual outbursts of EXO
2030+375. Figure~\ref{fig:asm} shows the 2-10 keV flux history for EXO 2030+375
measured with the {\em RXTE} ASM. The ASM routinely scans the sky with
$\sim 90$ second ``dwells" on each sky region. We averaged the ``dwells" over
4-day intervals to improve our sensitivity to EXO 2030+375 outbursts. Filled
circles in Figure~\ref{fig:asm} denote 4-day averages that are $\gtrsim 3
\sigma$ measurements. The majority of the apparent detections are near the time
of periastron passage denoted with dotted vertical lines.  In fact, for the 43 
periastron passages observed as of 2001 August with the {\em RXTE} ASM, since 
MJD 50135 (1996 February 21), only 2 fail to show $\gtrsim 3 \sigma$ evidence of
an outburst. The {\em RXTE} ASM saw evidence for an outburst in all but one of the
cases where no outburst was detected with BATSE (See Table~\ref{tab:outbursts}) 
in the period when both instruments were active. Prior to the launch of  
{\em RXTE}, BATSE occasionally failed to detect an outburst near EXO 2030+375's 
periastron passages, with the most consecutive missed outbursts being 3 
outbursts from MJD 49867-50032 (1995 May 30 - Nov 11). Combining the {\em RXTE}
and BATSE results, we find that EXO 2030+375 appears to have undergone an outburst every 
periastron passage for 82 periastron passages from 1991 April to 2001 August
(MJD 48361 - 52138). For the 11 periastron passages where no outburst was 
detected (9 prior to the launch of {\em RXTE}), EXO 2030+375 most likely still 
had an outburst that peaked just below our detection threshold. 

To investigate the orbital phasing of the outbursts, we determined the time of 
outburst peaks by fitting a Gaussian to the 1-day BATSE 20-50 keV pulsed flux 
measurements described in Section~\ref{sec:orb} for each outburst. In addition,
for each predicted periastron passage where {\em RXTE} ASM data were available, 
we fitted a Gaussian to 46.0214 days of single dwell 2-10 keV flux measurements
centered on the predicted periastron time.  Figure~\ref{fig:goao} shows the 
orbital phase of the outbursts versus time. Dashed lines indicate the intervals
of orbital phase where pulsations were detected with BATSE. Arrows
indicate the orbital phase range of outbursts where pulsations were detected 
with the {\em RXTE} PCA, the second of which was not detected with BATSE. 
Filled squares indicate the time of outburst peaks determined from BATSE pulsed
fluxes and open circles indicate peaks determined from ASM data. If the error on
the Gaussian centroid was larger than 5 days, that point was not plotted.
From MJD 48361-49900 (1991 Apr - 1995 Jul), the outbursts occurred at a stable 
orbital phase, peaking about 6 days after periastron passage. Three outbursts 
were not detected with BATSE following this interval, but when the outbursts 
were again detected after MJD 50000 (1995 Oct 10), they peaked about 4 days 
before periastron passage. The orbital phase of these outbursts slowly recovered
to about 2.5 days after periastron. 

\subsection{Optical/IR Observations\label{sec:oir}}

\begin{figure}
\plotone{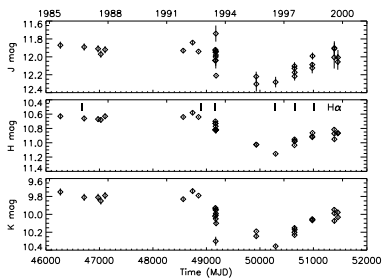}
\caption{Long-term IR history of the optical counterpart to EXO 2030+375. The
JHK magnitudes remain approximately constant from the giant outburst observed
with {\em EXOSAT} in 1985 until the BATSE observations (1991-2000). After about MJD 49000 (1993
January), the JHK magnitudes begin to decline, dropping to a minimum near MJD 
50300 (1996 August), followed by a slow brightening. Vertical lines near the top
of the center panel denote times of H$\alpha$ measurements.
\label{fig:hjk}}
\end{figure}

Infrared observations were obtained using the Continuously Variable Filter
Photometer on the 1.5m Carlos Sanchez Telescope at the Teide Observatory, 
Tenerife, Spain as part of the Southampton-Valencia monitoring campaign.
Figure~\ref{fig:hjk} shows long-term infrared photometric measurements of the 
optical counterpart to EXO 2030+375. Observations prior to MJD 49929 (1995 Jul
31) are taken
from Table~2 in \citet{Reig98}. For isolated Be stars, variations in these bands
are believed to be good indicators of the size of the Be star's equatorial disk.
However, when the Be star is in a binary system with a neutron star, the Be disk
is truncated at a resonance radius by tidal forces from the orbit of the neutron
star \citep{Okazaki01}. In these cases, as the disk cannot easily change size 
because of the truncation radius, changes in mass loss from the Be star produce
changes in the disk density, which can even become optically thick at IR 
wavelengths. In this case, the IR magnitudes and the H$\alpha$ equivalent width
are more related to the disk density than to the disk radius 
\citep[for example]{Negueruela01a, Miroshnichenko01}. Although there was a gap with no 
measurements from MJD 47106-48565 (1987 Nov - 1991 Nov), the magnitudes in all three bands remained at
the same level from near the end of the giant outburst observed with 
{\em EXOSAT} until about MJD 49000 (1993 Jan). From MJD 49000 until it reached 
a minimum near MJD 50300 (1996 Aug), the magnitudes in all 3 bands slowly 
became fainter, indicating a declining density in the Be disk. The Be star 
apparently did not go into a disk-loss phase as has been observed in other 
systems (e.g., A0535+262, Negueruela et al.\ 2000). Instead, the IR measurements
indicate that the density of the disk began to increase. 

One of the best direct tools for determining if there have been any
major structural changes in the circumstellar disk around the Be star
is the profile of the H$\alpha$ line. Presented in Figure~\ref{fig:ha} are a
sample of such profiles covering the period of interest; the details of
the observations are presented in Table~\ref{tab:ha}. The equivalent widths (EW)
of most of the observations have been published in \citet{Reig98}. In this paper
we are primarily concerned with the shapes of the H$\alpha$ profiles, none of 
which have previously been published.  

Looking at the 1997 and 1998 spectra one can see definite evidence
that the profile is not just a simple one, even though the signal to
noise ratio is rather lower than the earlier brighter spectra. The
extended nature of the profiles is indicative of a significantly
different circumstellar disk structure than before. Certainly the well
established correlation between the H$\alpha$ EW and the IR flux for
this source \citet{Reig98} enables one to use the IR flux as an
excellent indicator that changes were taking place around this time in
the disk.

Because of the weak flux in the profiles it is not possible to say
anything definite about the shapes of the profiles, but they appear
similar to the double structures seen in many such systems.  Certainly the
1997 and 1998 profiles are much wider than the 1996 one (all observed at
similar resolutions with the same telescope), indicating the presence of
material in the circumstellar disk moving at high velocities close to the
Be star.  This could well be direct evidence for the re-building of the
disk after the extended low period.

\citet{Parmar89} estimated a distance of 5 kpc to EXO 2030+375 and the fitted 
values from torque models in \citet{Reynolds96} are also in that range.  
However, optical data require a larger distance.\ \citet{Coe88} determined an 
interstellar extinction of $E(B-V) = 3.74$ from optical photometry which agreed
well with the X-ray column density determined with {\em EXOSAT}. Since EXO 
2030+375 is within 1.5\arcdeg\ of the galactic plane, we use the relationship 
between extinction and distance for objects in the plane \citep{Binney98} and
get $d = 7.1 \pm 0.2$ kpc (assuming an error of $\sim 0.1$ on the interstellar 
extinction).  

\section{Discussion}

\subsection{Improvements to Sensitivity and Orbital Parameters}

Improvements to our techniques used to search for pulsations in the BATSE data
reduced systematic errors to below statistical levels. These improvements
included using a modified $Z_n^2$ statistic which accounted for aperiodic noise
from either the measured source or others in the field of view such as Cygnus 
X-1 and automatically fitting Earth occultation steps from bright sources using
a database including source locations, dates of activity, and flux levels to 
determine which sources needed to be fitted. These improvements allowed us to
use much longer integrations for searches, e.g. 4 days, with accurately
determined errors. Searches of BATSE data using these new techniques resulted
in detection of 52 outbursts including several in the 2.5 year period from 1993
August to 1996 April when EXO 2030+375 was previously believed to be quiescent.
Our results show that EXO 2030+375 has undergone an outburst near most likely
every periastron passage for 9 years. From MJD 50643-51004 (1997 Jul - 1998 Jul),
EXO 2030+375 went undetected with BATSE for the longest interval, 7 periastron 
passages. However, {\em RXTE} PCA observations detected an outburst from EXO 
2030+375 on the third periastron passage missed with BATSE and the {\em RXTE} 
ASM detected outbursts for 6 of those periastron passages, suggesting that 
outbursts were still occurring but were below BATSE's detection threshold. 
Cygnus X-1 was noisy during this time interval. Additional noise plus a slight 
decrease in the intensity of EXO 2030+375 likely explains why BATSE missed 
these outbursts.

The 13 consecutive outbursts used by \citet{Stollberg99} to determine an orbit
for EXO 2030+375 provided good coverage from periastron passage to about 14 days
after periastron passage. Their orbit fitting was further complicated by the 
intrinsic spin frequency variations during these 13 brighter outbursts.  
Comparing panel (b) in Figure~\ref{fig:freqs} and Figure~\ref{fig:goao} shows where there were
problems with the \citet{Stollberg99} orbital parameters. The outbursts from 
MJD 50000-50700 (1995 Oct - 1997 Sep) covered earlier orbital phases, from about
7.5 days before until 6 days after periastron passage, than those included in 
the orbit fitting of \citet{Stollberg99}. These outbursts showed excess scatter in
the spin frequencies determined using the orbital parameters of 
\citet{Stollberg99}, indicating that fitting those outbursts would improve the 
orbital parameters. Fits to BATSE and {\em RXTE} data from these outbursts in 
addition to the outbursts fitted by \citet{Stollberg99} resulted in improved 
orbital parameters that removed the excess scatter in the spin frequencies. The
orbital parameters were further improved by including pulse frequencies 
determined from {\em EXOSAT} ME data during the initial giant outburst, which 
provided coverage of the entire orbit including two periastron passages with good
coverage of orbital phases from 21 to 7 days before periastron passage. Fits to
a broad range of orbital phases allowed us to decouple orbital and intrinsic
effects and to provide a good determination of the orbital parameters without
using complicated models for the intrinsic torque variations. Our new
orbital parameters are given in Table~\ref{tab:orb}.

\subsection{Orbital Phasing of Outbursts}

The outbursts prior to MJD 50000 (1995 Oct) peaked at a very regular orbital phase of about
6 days after periastron passage (see Figure~\ref{fig:goao}). The outburst just
after MJD 50000 peaked at a much earlier orbital phase, 4 days before periastron
passage and then gradually recovered to peak at a new stable orbital phase of 
about 2.5 days after periastron passage. A possible explanation of the shift in
orbital phase of the outbursts is a density perturbation (global one armed 
oscillation) in the Be disk. Evidence for these density perturbations is seen in
the H$\alpha$ line profiles for several Be/X-ray binaries 
\citep[for example]{Negueruela01a, Negueruela01b}. When a density perturbation is present, 
the H$\alpha$ line is double peaked. The relative size of the two peaks changes
with a cycle of several years. The density perturbation produces a non-axially 
symmetric Be disk. If the pulsar interacts with a region of the disk affected by
the perturbation, more material would be available for accretion, possibly 
causing a shift in outburst phase. The trend in the orbital phases from MJD 
50000-50600 (1995 Oct - 1997 Jun) in Figure~\ref{fig:goao} has a slope of 
$m = 0.0085 \pm 0.0017$. This slope can be expressed in terms of a beat 
frequency between the orbital period and some other period. Assuming the other 
period is longer than the orbital period, as is expected for global one-armed 
oscillations, then 
\begin{equation}
m = \frac{\nu_{\rm beat}}{\nu_{\rm orb}}-1
\end{equation}
where $\nu_{\rm beat}$ is the beat frequency and $\nu_{\rm orb}$ is the orbital
frequency. If this trend is due to beating between the orbital period and the 
period of the density perturbation, then the beat frequency of is given by 
\begin{equation}
\nu_{\rm beat} = \nu_{\rm orb} + \nu_{\rm perturb} 
\end{equation}
where $\nu_{\rm perturb}$ is the frequency of the density perturbation. Solving
for $\nu_{\rm perturb}$, we get a perturbation frequency of $(1.8 \pm 0.4) \times
10^{-4}$ cycles day$^{-1}$ or equivalently, a period of $15 \pm 3$
years for the density perturbation to propagate around the Be disk. No significant 
change in X-ray intensity was seen when the outbursts shifted in orbital phase,
although 3 outbursts went undetected near the time when the shift occurred. 
{\em EXOSAT} observations of the second outburst of EXO 2030+375
after its discovery (1985 October 29-November 3, MJD 46367-46372) show that the
outburst peaked at $9.5 \pm 1.1$ days after periastron, assuming that the peak 
flux detected with {\em EXOSAT} is the peak of a normal outburst. Interestingly,
this is 3.5 days later than the peak time observed in the pre-MJD 50000 (1995
Oct) BATSE data, which is also 3.5 days later than the peak time observed after
MJD 51000 (1998 Jul), when the peak time had stopped changing rapidly and
occurred about 2.5 days after periastron. This suggests that another shift in 
outburst phase may have occurred between the {\em EXOSAT} and BATSE 
observations. However, if such a shift occurred, it suggests a shorter 
propagation period of $\lesssim 10$ years. 

Optical observations of the H$\alpha$ profile (Figure~\ref{fig:ha}) clearly 
indicate that the structure of the circumstellar disk around the Be star changed
significantly at some time between the 1996 July (MJD 50273) observation and
the 1997 August (MJD 50661) observation. These observations support the idea of
a global one-armed oscillation propagating in the Be disk suggested by the shift
in orbital phase of the outbursts. However, sparse observations of the H$\alpha$
profile do not allow us to directly correlate H$\alpha$ profile changes with
changes in the X-ray outbursts.

\subsection{Relationship between X-ray and IR Measurements\label{sec:xir}}

\begin{figure}
\plotone{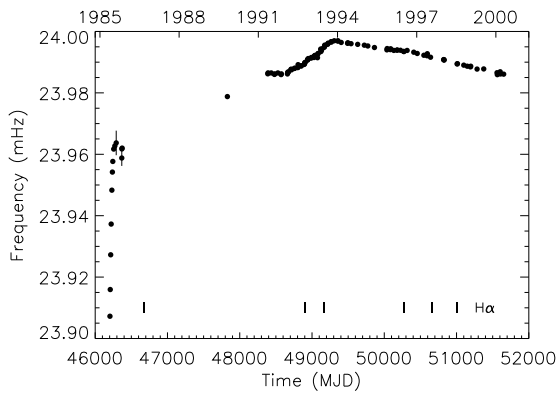}
\caption{Long-term spin frequency history for EXO 2030+375. All frequencies are
barycentered and corrected for the pulsar's orbital motion using the parameters
in row 5 of Table~\ref{tab:orb}. Vertical lines near the bottom of the plot
denote times of H$\alpha$ measurements.
\label{fig:longterm}}
\end{figure}

Comparing the IR measurements in Figure~\ref{fig:hjk} to the X-ray 
measurements (Figures~\ref{fig:freqs} and \ref{fig:longterm}) suggests a 
relationship between the IR behavior and the X-ray activity. The IR
measurements indicate that the Be disk was fairly stable and roughly constant
in density from near the end of the initial giant outburst until MJD 49000
(1993 Jan). Figure~\ref{fig:longterm} indicates that the pulsar was spinning up
for most of this period. Near MJD 49000, the density of the disk began to
decline. Since the disk was becoming less dense, the reservoir of material
available to the pulsar near periastron passage was slowly reduced. After MJD
49250 (1993 Sep), the  X-ray pulsed flux responded to the lower density disk
and dropped dramatically. Peak 20-50 keV pulsed fluxes dropped from $(3-5)
\times 10^{-10}$ \ergcms\ to  $\lesssim 1.5 \times 10^{-10}$ \ergcms. The global
spin-up rate of the pulsar took longer to respond. After about MJD 49250, the
spin-up rate slowed and by MJD  49400 (1994 Feb), the pulsar had begun to
spin-down. No obvious response in the X-rays was seen to the slow increase in
density of the Be disk indicated by the IR measurements after MJD 50300 (1996
Aug). Perhaps the disk had not yet become dense enough to increase the mass
accretion rate to a level where the pulsar would begin to spin-up. 

To determine whether or not the observed spin-down was likely due to centrifugal
inhibition of accretion \citep{Stella86}, i.e., the propeller effect
\citep{Illarionov75}, we estimate the flux at the onset of this effect by
equating the magnetospheric radius to the corotation radius. The magnetospheric
radius is given by \citep{Pringle72, Lamb73}
\begin{equation}
r_{\rm m} \simeq k (G M)^{1/7} \mu^{-2/7} L^{-2/7} R^{-2/7}\label{eqn:rm} 
\end{equation}
where $G$ is the gravitational constant; $M$ and $R$ are the mass and radius of
the neutron star; and $L$ is the luminosity. $k$ is a constant factor of order
1. Equation~\ref{eqn:rm} with $k \simeq 0.91$ gives the Alfv\'en radius for 
spherical accretion and with $k \simeq 0.47$ gives the magnetospheric radius 
derived by \citet{Ghosh79}. The corotation radius is given by 
\begin{equation}
r_{\rm co} = (G M)^{1/3} (2 \pi \nu)^{-2/3}\label{eqn:rco}
\end{equation}
where $\nu$ is the spin frequency of the pulsar. Setting $r_{\rm m} = r_{\rm
co}$ gives the threshold flux for the onset of centrifugal inhibition of
accretion, i.e.,
\begin{equation}
F_{\rm x}^{\rm min} \simeq 3 \times 10^{-10}\ {\rm ergs}\ {\rm cm}^{-2}\
{\rm s}^{-1}\ k^{7/2} \mu_{30}^2 M_{1.4}^{-2/3} R_6^{-1} P_{\rm 41.7s}^{-7/3} 
d_{\rm kpc}^{-2}\label{eqn:cia}
\end{equation}
where $\mu_{30}$, $M_{1.4}$, $R_6$, and $P_{\rm 41.7 s}$ are the pulsar's 
magnetic moment in units of $10^{30}$ G cm$^{3}$, mass in units of 1.4 
$M_{\odot}$, radius in units of $10^6$ cm, and spin period in units of 41.7 
seconds, respectively. \citet{Reynolds96} fitted 3 accretion torque models, a
simple spherical accretion model, the \citet{Ghosh79} model, and the
\citet{Wang87} model to the giant outburst to estimate values for $\mu_{30}$ and
$d_{\rm kpc}$. For the simple spherical accretion model,
where $k \simeq 0.91$, \citet{Reynolds96} obtained $\mu_{30} \simeq 5$ for an
assumed distance of $d_{\rm kpc} = 5$. Both the \citet{Ghosh79} model and the 
\citet{Wang87} model use $k \simeq 0.47$. \citet{Reynolds96} obtained 
$\mu_{30} \simeq 12$ and $d_{\rm kpc} \simeq 5.2$ from fits to the
\citep{Ghosh79} model. \citet{Parmar89} also fitted the \citet{Ghosh79} model, 
obtaining $\mu_{30} \simeq 20$ and $d_{\rm kpc} \simeq 5.3$ for the first nine period measurements and $\mu_{30} \simeq 11$ and
$d_{\rm kpc} \simeq 5.0$ for the first 10 period measurements.
\citet{Reynolds96} fit of the \citet{Wang87} model yielded the lowest distance $
d_{\rm kpc} \simeq 4.1$ with $\mu_{30} \simeq 10$.  Substituting
\citet{Reynolds96} values yields $F_{\rm x}^{\rm min} \simeq (1.1-2.2) \times
10^{-10}$ \ergcms\ while \citet{Parmar89}'s values yield $F_{\rm x}^{\rm min} 
\simeq (1-3) \times 10^{-10}$ \ergcms.  

The minimum flux where pulsations were first detected during the faint 1996
July outburst observed with {\em RXTE} was $3.3
\times 10^{-10}$ \ergcms\  \citep[2.7-30 keV]{Reig99}, which is comparable to
the lowest flux of $4 \times 10^{-10}$ \ergcms\ observed by \citet{Parmar89} 
before pulsations became undetectable in the giant outburst. Since pulsations 
were detected with {\em RXTE} and no significant changes in the 2-10 keV pulse profile 
were observed relative to higher fluxes \citep{ReigCoe98}, this flux is most 
likely above the threshold for centrifugal inhibition of accretion, which is 
consistent with our calculations in the previous paragraph. On 1985 August 25,
when pulsations were not detected, \citet{Parmar89} measured an upper limit of 
$1.3 \times 10^{-11}$ \ergcms\ on the flux from EXO 2030+375, which places a
lower bound on $F_{\rm x}^{\rm min}$, that is also consistent with our
calculations, assuming that the observed sudden drop in flux was due to
centrifugal inhibition of accretion. 
 
\subsection{Spin-up vs. flux correlations}

In giant outbursts of Be/X-ray binaries, accretion disks are expected to be
present and indeed, evidence for an accretion disk, based on a correlation 
between the observed flux and spin-up rate, has been found for several sources
including EXO 2030+375 \citep{Parmar89, Reynolds96, Stollberg99} during 
giant outbursts of Be/X-ray binaries \citep{Wilson98, Bildsten97}. Independent
evidence for an accretion disk based on the detection of quasi-periodic
oscillations during a giant outburst has been found for EXO 2030+375
\citep{Angelini89} and A0535+262 \citep{Finger96b}. In addition to the expected
correlation between observed flux and spin-up rate in the giant outburst of EXO
2030+375, the BATSE data suggest a correlation is also present in the brighter
normal outbursts of this system (See Figure~\ref{fig:fdotvsflux}). Until 
recently, normal outbursts were believed to be due to direct wind accretion from
the Be disk, so significant spin-up was not expected because wind accretion is 
not believed to be very efficient at transferring angular momentum \citep{Ruffert97}. However, evidence for spin-up 
during normal outbursts has been observed in GS 0834--430 \citep{Wilson97}, 2S 
1417--624 \citep{Finger96a}, 2S 1845--024 \citep{Finger99}, and
previously in EXO 2030+375 \citep{Stollberg99}. In BATSE observations of 2S
1845--024, a correlation between the spin-up rate and the pulsed flux was also
observed. 

The comparison of the BATSE and {\rm EXOSAT} data in
Figure~\ref{fig:fdotvsflux}, despite an uncertain bolometric correction,
shows that at the lower fluxes measured with BATSE the correlation falls off 
more rapidly than a power law. This suggests either that spin-down torques 
become important or perhaps that a disk forms during the outburst and we are 
averaging over periods of wind and disk accretion. The data are clearly 
inconsistent with the power law index of $6/7$ predicted from simple accretion 
theory, which does not consider spin-down torques. The \citet{Ghosh79} model 
fitted by \citet{Reynolds96}, which assumes an accretion disk is present, 
roughly follows the trend observed in the data; however, the brightest 
{\em EXOSAT} observations, which drive the fit, clearly deviate from the model.
The observed flux is given by $F = (\beta L)/(4 \pi d^{2})$, where $\beta$ 
is a beaming factor.  This beaming factor, which is typically assumed to be
equal to one for simplicity, depends on the pattern of emitted radiation at the
pulsar and cannot be determined without modeling of the pulse 
profiles (which is beyond the scope of this work). Because large luminosity 
dependent 1-10 keV pulse profile variations were observed during the giant outburst 
\citep{PWS89}, one would expect that the beaming factor was also changing with 
luminosity. To fit any of the discussed models, the beaming factor must decrease
with increasing luminosity. 

The spin-up rate and its correlation with pulsed flux during the earlier 
outbursts of EXO 2030+375 observed with BATSE suggest an accretion disk may be 
present. A disk will form if the specific angular momentum of the material
accreted from the Be star's disk is comparable to the Keplerian specific angular
momentum at the magnetospheric radius. The specific angular momentum $\ell$ of 
the accreted material is given by
\begin{equation} 
\ell = 2 \pi I \dot \nu \dot M^{-1} \simeq (4.8-9.1) \times 10^{16}\ \rm{cm}\ 
\rm{s}^{-1}\ d_{\rm 7.1 kpc}^{-2},\label{eqn:ell}
\end{equation}
assuming $\dot M = L (G M/ R)^{-1}$. Here $\dot \nu = 6.5 \times 10^{-13}$ Hz 
s$^{-1}$ is the spin-up rate, $\dot M$ is the mass accretion rate, $L =
(0.8-1.6) \times 10^{37}$ \ergss\ $d_{\rm 7.1 kpc}^2$ is the luminosity, and 
$d_{\rm 7.1 kpc}$ is the distance in units of 7.1 kiloparsecs. The spin-up rate
and luminosity are outburst averaged values from a typical bright outburst 
observed with BATSE in 1992 February. The luminosity is estimated from BATSE
pulsed flux of $2.5 \times 10^{-10}$ \ergcms\ using a bolometric correction of 
$8.0 \pm 2.4$ (See Section~\ref{sec:torq}), and a distance of 7.1 kpc (See
Section~\ref{sec:oir}). We have assumed typical pulsar parameters, listed in 
Section~\ref{sec:xir}. If $\ell \sim \ell_{\rm m} = (GMr_{\rm m})^{1/2}$, the 
Keplerian specific angular momentum at the magnetospheric radius, a disk will 
form. Once a disk has formed the specific angular momentum of accreting material
is maintained near $\ell_{\rm m}$. At $F_{\rm x}^{\rm min}$ 
(Equation~\ref{eqn:cia}), $r_{\rm m} = r_{\rm co}$. Hence $\ell_{\rm m}$ can be 
expressed as a function of $F_{\rm x}^{\rm min}$ and $r_{\rm co}$, which is 
independent of the distance and magnetic field of the pulsar.
\begin{equation}
\ell_{\rm m} \simeq 6.1 \times 10^{17}\ {\rm cm}^2\ {\rm s}^{-1}\ 
P_{\rm 41.7s}^{1/3} \left(\frac{F}{F_{\rm x}^{\rm min}}\right)^{-1/7}
\end{equation} 
In the previous section, using {\em RXTE} and {\em EXOSAT} observations,
we determined $1.3 \times 10^{-11}$ \ergcms\ $\lesssim F_{\rm x}^{\rm min}
\lesssim 3.3 \times 10^{-10}$ \ergcms.  With the estimated average 
bolometric flux of $(1.4-2.6) \times 10^{-9}$ \ergcms\ from the 1992 February 
outburst used in Equation~\ref{eqn:ell}, this yields $\ell_{\rm m} \simeq 
(2.9-5.0) \times 10^{17}$ cm$^2$ s$^{-1}$, which is within an order of magnitude of $\ell
\simeq (4.8-9.1) \times 10^{16}$ cm$^2$ s$^{-1}$, suggesting a disk is likely present 
because considerable angular momentum is present in the system. If a disk forms
during the outburst, our average values of $\dot \nu$ and $F$ used in these 
calculations would include periods of wind accretion and periods of disk 
accretion, possibly explaining why $\ell \simeq (0.1-0.3) \ell_{\rm m}$ rather 
than being a larger fraction. In contrast, for the wind-fed system Vela 
X-1 where a disk is not expected to be present, $\dot \nu \simeq 6 \times 
10^{-14}$ Hz s$^{-1}$ \citep{Inam00}, $L \simeq 2 \times 10^{38}$ \ergss, and 
$\mu \simeq 2.1 \times 10^{30}$ G cm$^{3}$ \citep{Makishima99} leading to 
$\ell \simeq 3.5 \times 10^{14}$ cm$^2$ s$^{-1}$ and $\ell_{\rm m} \simeq 2.2 
\times 10^{17}$ cm$^2$ s$^{-1}$, i.e., $\ell \simeq 0.002 \ell_{\rm m}$. 

\subsection{Current Models}

Although the direct wind accretion model has difficulty explaining 
evidence for an accretion disk during normal outbursts, the viscous decretion
disk model \citep{Lee91, Porter99, Okazaki01a} provides a natural explanation.
In this model, the inner edge of the Be disk has a Keplerian velocity. Viscosity
conducts material outwards, so that it moves in quasi-Keplerian orbits with low
radial velocities. The radial outflow is subsonic for the orbital sizes of all
Be/X-ray binaries with a known solution. This model successfully accounts for
most observations of Be disks \citep{Okazaki01}. \citet{Negueruela01a} found
that tidal interaction of the neutron star truncates the circumstellar Be disk.
In the disk, truncation occurs when the outward viscous torque is less than the
inward resonant torque which truncates the disk at a resonant radius. Because of
the truncation, the Be disk cannot reach a steady state \citep{Okazaki01}.
According to the modeling of \citet{Okazaki01}, the Be disk in EXO 2030+375 is 
likely truncated at the 4:1 resonance radius, which is close to the radius of
the critical lobe at periastron. If the truncation radius is close to or 
slightly beyond the critical lobe radius at periastron, material with high 
angular momentum will flow through the first Lagrangian point to the neutron 
star, making formation of a transient accretion disk likely \citep{Okazaki01} 
in a normal outburst. Hence this model provides a reasonable explanation of the
brighter outbursts. 

\begin{figure}
\plotone{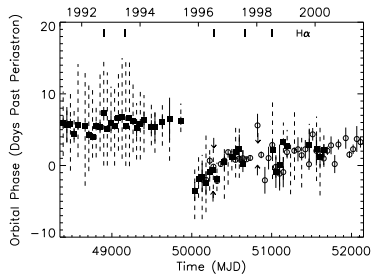}
\caption{Orbital phase of EXO 2030+375 outburst peaks versus time from 1991
April - 2001 August. Dashed lines indicate the orbital phases when pulsations from EXO 2030+375 were detected with
BATSE. Filled squares indicate the times of outburst peaks estimated from
Gaussian fits to 1-day BATSE pulsed fluxes corresponding to BATSE pulsed phase
measurements. Open circles indicate the times of outburst peaks estimated from
Gaussian fits to 46.0214 days of single dwell {\em RXTE} ASM 2-10 keV flux 
measurements centered on the periastron epoch for each outburst. Arrows denote
the orbital phase range of {\em RXTE} PCA detections. Vertical lines
near the top of the plot denote times of H$\alpha$ measurements.
\label{fig:goao}}
\end{figure}

\begin{figure}
\resizebox{\textwidth}{!}{
\includegraphics[angle=180]{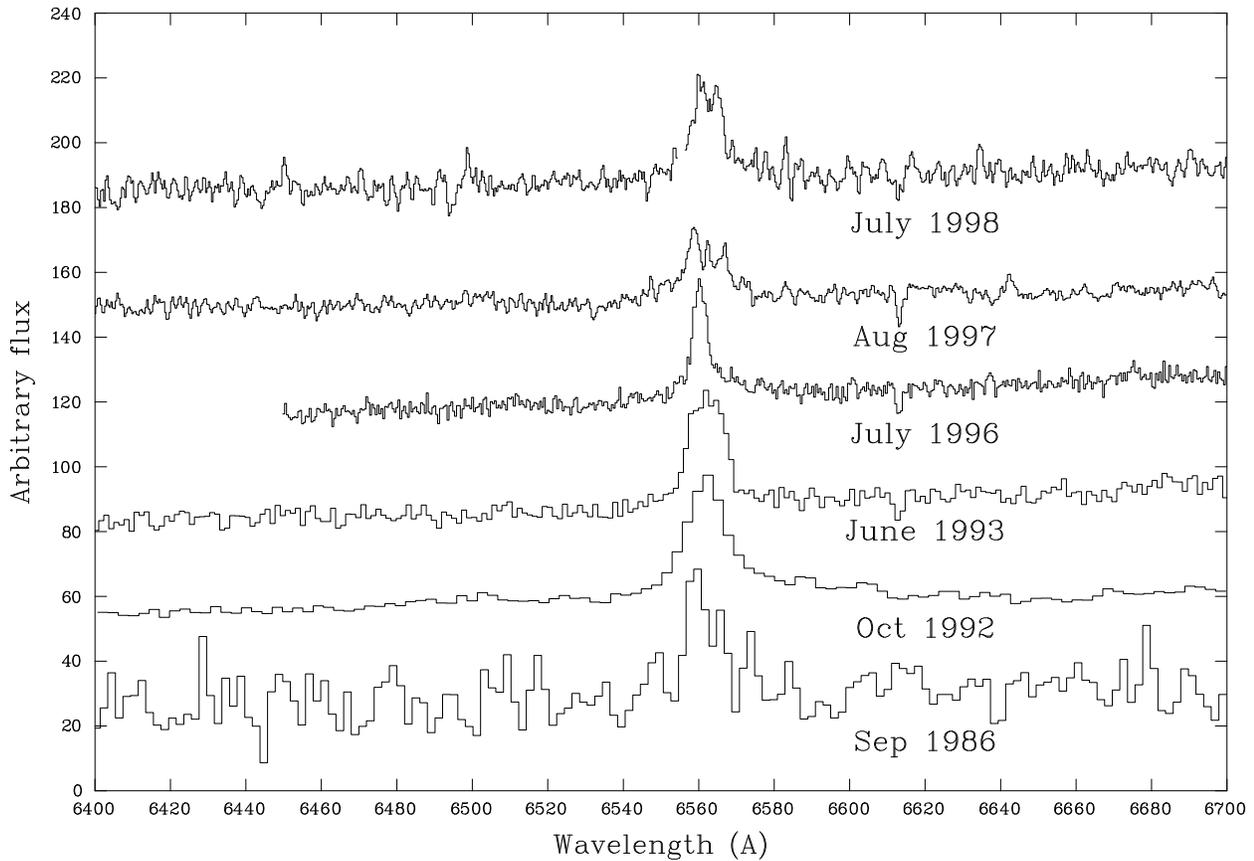}}
\caption{$H\alpha$ line profiles from 6 observations of EXO 2030+375. Times of 
the H$\alpha$ profiles are denoted with vertical bars in Figures~\ref{fig:freqs}, 
\ref{fig:longterm}, \ref{fig:goao}, and \ref{fig:hjk}. The extended nature 
of the 1997 (MJD 50661) and 1998 (MJD 51009) profiles suggests a different 
circumstellar disk structure than before. The 1996 (MJD 50273), 1997, and 1998 
data were taken with almost the same instrumental configuration and hence are 
directly comparable. The earlier data were taken with poorer wavelength 
resolution. (See Table~\ref{tab:ha}.) 
\label{fig:ha}}
\end{figure}

We propose the following scenario to explain the IR and X-ray observations. 
The Be disk is truncated at a 4:1 resonance radius. Following \citet{Okazaki01},
there should be an outburst at every periastron unless the circumstellar disk
disappears. Around MJD 49000 (sometime in 1993), a major structural change
occurred in the Be star's circumstellar disk. It became much less dense, as
shown by the change in IR magnitudes and the H$\alpha$ equivalent width. Much
less matter was available for accretion, and as a consequence, the X-ray flux
dropped and the neutron star spin-up ended. At the same time, or shortly
afterward, a density wave (or global one-armed oscillation) developed and began
to precess, without interacting with the neutron star's orbit. Around MJD 50000
(late in 1995), the density perturbation interacted with the neutron star's
orbit, at a phase corresponding to about 3 days prior to periastron passage,
producing an X-ray outburst peaked at that phase. This implies that the
precession of the density perturbation was prograde, in the same sense as the 
neutron star's orbital motion. Sometime after about MJD 50600 (mid-1997), the 
density perturbation lost contact with the neutron star's orbit, in a position
symmetrical with respect to periastron, i.e., about 3 days after periastron.
This ended the fast migration of the outburst peaks in orbital phase. Some slow
migration may still be present in Figure~\ref{fig:goao}, which may eventually
lead to the previous value of 6 days after periastron, after a complete
precession period of the perturbed disk.

\acknowledgements
This research has made use of data obtained from the High Energy Astrophysics 
Science Archive Research Center (HEASARC), provided by NASA's Goddard Space 
Flight Center (GSFC). {\em RXTE} ASM quick-look results were provided by the ASM/{\em
{\em RXTE}} teams at MIT and at the GSFC SOF and GOF. We are grateful to the
support staff of the Telescopio Carlos Sanchez and the Service Programme of the
Isaac Newton Group for help obtaining much of the optical and IR data. The TCS
is operated on the Teide Observatory by the Instituto de Astrofisica de
Canarias. We thank an anonymous referee whose suggestions improved our paper.

\clearpage

\begin{deluxetable}{lccll}
\tabletypesize{\scriptsize}
\tablecaption{EXO 2030+375 Outburst Detections}
\tablewidth{0pt}
\tablehead{
\colhead{Number\tablenotemark{a}} & \colhead{Start Date} & 
\colhead{End Date} & \colhead{Start Date} & \colhead{Stop Date} \\
 & \colhead{(MJD)} & \colhead{(MJD)} & \colhead{(Calendar)} & 
 \colhead{(Calendar)}}
\startdata
 \nodata\tablenotemark{b}  & 46203 & 46303 & 1985 May 18 & 1985 Aug 26 \\
 \nodata\tablenotemark{b}  & 46367 & 46373 & 1985 Oct 29 & 1985 Nov 4 \\
 \nodata\tablenotemark{c} & 47828 & 47830 & 1989 Oct 29 & 1989 Oct 31 \\
  1 & 48385 & 48395 & 1991 May  9 & 1991 May 19 \\
  2 & 48433 & 48439 & 1991 Jun 26 & 1991 Jul 2 \\
  3 & 48475 & 48487 & 1991 Aug  7 & 1991 Aug 19 \\
  4 & 48524 & 48531 & 1991 Sep 25 & 1991 Oct 2 \\
  5 & 48569 & 48583 & 1991 Nov  9 & 1991 Nov 23 \\
  7\tablenotemark{d}  & 48659 & 48673 & 1992 Feb  7 & 1992 Feb 21 \\
  8\tablenotemark{d}  & 48705 & 48718 & 1992 Mar 24 & 1992 Apr 6 \\
  9\tablenotemark{d}  & 48753 & 48764 & 1992 May 11 & 1992 May 22 \\
 10\tablenotemark{d}  & 48799 & 48807 & 1992 Jun 26 & 1992 Jul 4 \\
 11\tablenotemark{d}  & 48846 & 48858 & 1992 Aug 12 & 1992 Aug 24 \\
 12\tablenotemark{d}  & 48890 & 48906 & 1992 Sep 25 & 1992 Oct 11 \\
 13\tablenotemark{d}  & 48936 & 48950 & 1992 Nov 10 & 1992 Nov 24 \\
 14\tablenotemark{d}  & 48984 & 48994 & 1992 Dec 28 & 1993 Jan  7 \\
 15\tablenotemark{d}  & 49028 & 49042 & 1993 Feb 10 & 1993 Feb 24 \\
 16\tablenotemark{d}  & 49076 & 49088 & 1993 Mar 30 & 1993 Apr 11 \\
 17\tablenotemark{d} & 49120 & 49137 & 1993 May 13 & 1993 May 30 \\
 18\tablenotemark{d}  & 49166 & 49182 & 1993 Jun 28 & 1993 Jul 14 \\
 19\tablenotemark{d}  & 49214 & 49228 & 1993 Aug 15 & 1993 Aug 29 \\
 20 & 49261 & 49271 & 1993 Oct  1 & 1993 Oct 11 \\
 21 & 49305 & 49314 & 1993 Nov 14 & 1993 Nov 23 \\
 22 & 49351 & 49362 & 1993 Dec 30 & 1994 Jan 10 \\
 23 & 49398 & 49407 & 1994 Feb 15 & 1994 Feb 24 \\
 25 & 49491 & 49498 & 1994 May 19 & 1994 May 26 \\
 26 & 49536 & 49543 & 1994 Jul  3 & 1994 Jul 10 \\
 28 & 49629 & 49636 & 1994 Oct  4 & 1994 Oct 11 \\
 30 & 49719 & 49726 & 1995 Jan  2 & 1995 Jan  9 \\
 33 & 49861 & 49866 & 1995 May 24 & 1995 May 29 \\
 37 & 50033 & 50041 & 1995 Nov 12 & 1995 Nov 20 \\
 38 & 50082 & 50089 & 1995 Dec 31 & 1996 Jan  7 \\
 39 & 50125 & 50135 & 1996 Feb 12 & 1996 Feb 22 \\
 40 & 50176 & 50183 & 1996 Apr  3 & 1996 Apr 10 \\
 41 & 50221 & 50228 & 1996 May 18 & 1996 May 25 \\
 42\tablenotemark{e,f} & 50265 & 50275 & 1996 Jul  1 & 1996 Jul 11 \\
 43 & 50313 & 50318 & 1996 Aug 18 & 1996 Aug 23 \\
 44\tablenotemark{g} & 50359 & 50367 & 1996 Oct  3 & 1996 Oct 11 \\
 45 & 50404 & 50412 & 1996 Nov 17 & 1996 Nov 25 \\
 46 & 50452 & 50458 & 1997 Jan  4 & 1997 Jan 10 \\
 47 & 50500 & 50504 & 1997 Feb 21 & 1997 Feb 25 \\
 48 & 50545 & 50553 & 1997 Apr  7 & 1997 Apr 15 \\
 49 & 50591 & 50599 & 1997 May 23 & 1997 May 31 \\
 50 & 50639 & 50642 & 1997 Jul 10 & 1997 Jul 13 \\
 51\tablenotemark{g} & 50684 & 50692 & 1997 Aug 24 & 1997 Sep  1 \\
 52\tablenotemark{g} & 50728 & 50735 & 1997 Oct  7 & 1997 Oct 14 \\
 54\tablenotemark{e} & 50821 & 50828 & 1998 Jan  8 & 1998 Jan 15 \\
 55\tablenotemark{g} & 50867 & 50876 & 1998 Feb 23 & 1998 Mar  4 \\
 56\tablenotemark{g} & 50915 & 50919 & 1998 Apr 12 & 1998 Apr 16 \\
 57\tablenotemark{g} & 50956 & 50963 & 1998 May 23 & 1998 May 30 \\
 58 & 51005 & 51017 & 1998 Jul 11 & 1998 Jul 23 \\
 59 & 51051 & 51056 & 1998 Aug 26 & 1998 Aug 31 \\
 60 & 51096 & 51102 & 1998 Oct 10 & 1998 Oct 16 \\
 61 & 51143 & 51152 & 1998 Nov 26 & 1998 Dec  5 \\
 62 & 51189 & 51197 & 1999 Jan 11 & 1999 Jan 19 \\
 63 & 51233 & 51241 & 1999 Feb 24 & 1999 Mar  4 \\
 64 & 51281 & 51292 & 1999 Apr 13 & 1999 Apr 24 \\
 65\tablenotemark{g} & 51327 & 51339 & 1999 May 29 & 1999 Jun 10 \\
 66 & 51372 & 51380 & 1999 Jul 13 & 1999 Jul 21 \\
 67\tablenotemark{g} & 51415 & 51428 & 1999 Aug 25 & 1999 Sep  7 \\
 68 & 51469 & 51474 & 1999 Oct 18 & 1999 Oct 23 \\
 69\tablenotemark{g} & 51511 & 51515 & 1999 Nov 29 & 1999 Dec  3 \\
 70 & 51558 & 51567 & 2000 Jan 15 & 2000 Jan 24 \\
 71 & 51601 & 51612 & 2000 Feb 27 & 2000 Mar  9 \\
 72 & 51651 & 51657 & 2000 Apr 17 & 2000 Apr 23 \\
 73\tablenotemark{g} & 51691 & 51707 & 2000 May 27 & 2000 Jun 12 \\
 74\tablenotemark{g} & 51740 & 51751 & 2000 Jul 15 & 2000 Jul 26 \\
 75\tablenotemark{g} & 51775 & 51795 & 2000 Aug 19 & 2000 Sep  8 \\
 76\tablenotemark{g} & 51835 & 51847 & 2000 Oct 18 & 2000 Oct 30 \\
 78\tablenotemark{g} & 51927 & 51935 & 2001 Jan 18 & 2001 Jan 26 \\
 79\tablenotemark{g} & 51972 & 51976 & 2001 Mar  4 & 2001 Mar  8 \\
 80\tablenotemark{g} & 52015 & 52028 & 2001 Apr 16 & 2001 Apr 29 \\
 81\tablenotemark{g} & 52063 & 52072 & 2001 Jun  3 & 2001 Jun 12 \\
 82\tablenotemark{g} & 52111 & 52123 & 2001 Jul 21 & 2001 Aug  2 \\
\enddata
\tablenotetext{a}{Periastron passage number since {\em CGRO} launch.}
\tablenotetext{b}{Outburst dates from {\em EXOSAT} detections.}
\tablenotetext{c}{Outburst dates from {\em Ginga} detections.}
\tablenotetext{d}{Included in orbit fit of
\protect\citet{Stollberg99}}
\tablenotetext{e}{Outburst dates from {\em RXTE} PCA detections.}
\tablenotetext{f}{{\em RXTE} PCA detected outburst used in all orbit fits.}
\tablenotetext{g}{$\gtrsim 3 \sigma$ detections with the {\em RXTE} ASM. Not
used in orbit fitting.} 
\label{tab:outbursts}
\end{deluxetable}

\begin{deluxetable}{lllllll}
\tabletypesize{\scriptsize}
\tablecaption{EXO 2030+375 Orbit Fits}
\tablewidth{0pt}
\tablehead{
 \colhead{Fit\tablenotemark{a}} & \colhead{$P_{\rm orb}$} & \colhead{$T_{\rm peri}$} &
  \colhead{$a_x \sin i$} & \colhead{e} & \colhead{$\omega$} & 
  \colhead{$\chi^2/$dof} \\
  & \colhead{(days)} & \colhead{(JD)} & \colhead{(lt-sec)} & & & }   
\startdata
S99\tablenotemark{b}  & $46.02 \pm 0.02$ & $2448937.0 \pm 0.2$ 
 & $262 \pm 24$ & $0.37 \pm 0.02$ & $223\arcdeg.5 \pm 4\arcdeg.3$ 
 & $103.31/102$ \\
R only & 46.02(fixed) & $2450271.5 \pm 0.2$ & $236.7 \pm 6.2$ 
 & $0.416 \pm 0.004$ & $206\arcdeg.9 \pm 6\arcdeg.3$ & $15.25/16$ \\
R+13 B & $46.026 \pm 0.003$ & $2450179.56 \pm 0.02$
 & $241.5 \pm 3.9$ & $0.414 \pm 0.004$ & $212\arcdeg.0 \pm 0\arcdeg.6$
 & $174.9/147$ \\
2 R+51 B & $46.023 \pm 0.001$ & $2450317.61 \pm 0.02$ 
 & $241.4 \pm 3.5$ & $0.413 \pm 0.003$ & $211\arcdeg.1 \pm 0\arcdeg.6$ 
 & $549.0/333$ \\
E +2 R +51B & $46.0214 \pm 0.0005$ 
 & $2450547.72 \pm 0.02$ & $235.8 \pm 1.8$ & $0.419 \pm 0.002$ 
 & $211\arcdeg.2 \pm 0\arcdeg.6$ & $556.3/338$ \\
\enddata
\tablenotetext{a}{Fit column designates number of outbursts used from each of
3 instruments, B=BATSE, E={\em EXOSAT}, and R={\em RXTE} for fits in this 
paper.}
\tablenotetext{b}{Orbital parameters from \protect\citet{Stollberg99}.}
\label{tab:orb}
\end{deluxetable}

\begin{deluxetable}{lllcc}
\tabletypesize{\scriptsize}
\tablecaption{Details of Spectroscopic Observations\label{tab:ha}}
\tablewidth{0pt}
\tablehead{
\colhead{Date} & \colhead{MJD} & \colhead{Telescope} & 
 \colhead{Spectral} & \colhead{$H\alpha$ Equivalent} \\
  & & & \colhead{Resolution} & \colhead{Width} \\
  & & & \colhead{(\AA/pixel)} & \colhead{(\AA)} }
\startdata
7 Sep 1986     & 46680  & INT		& 2.03		& -15.0		\\
2 Oct 1992     & 48897  & WHT		& 2.72		& -20.2		\\
29 Jun 1993   & 49167 	& WHT		& 1.33		& -18.0		\\
9 Jul 1996    & 50273	& WHT		& 0.50		& -6.8		\\
1 Aug 1997     & 50661	& WHT		& 0.40		& -5.8		\\
15 Jul 1998   & 51009	& WHT		& 0.40		& -8.0		\\
\enddata
\tablecomments{The telescopes referred to are : INT - Isaac Newton Telescope, WHT 
- William Herschel Telescope. Both telescopes are located on the island of La 
Palma, Spain.}
\end{deluxetable} 


\clearpage
\onecolumn
\end{document}